\documentstyle[epsfig,aps]{revtex} 

\newcommand{\gb}{\beta}

\newcommand{\gve}{\varepsilon}

\newcommand{\gD}{\Delta}

\tightenlines
\begin{document}


\twocolumn[
\hsize\textwidth\columnwidth\hsize\csname @twocolumnfalse\endcsname


\title{In-medium nucleon-nucleon potentials in configuration space}
\author{M. Beyer$^*$, S. A. Sofianos}
\address{Physics Department, University of South Africa
         0003, Pretoria, South Africa}
\date{\today}                     
\maketitle
\begin{abstract} 
  Based on the thermodynamic Green function approach two-nucleon
  correlations in nuclear matter at finite temperatures are revisited.
  To this end, we derive phase equivalent effective $r$-space
  potentials that include the effect of the Pauli blocking at a given
  temperature and density.  These potentials enter into a
  Schr\"odinger equation that is the $r$-space representation of the
  Galitskii-Feynman equation for two nucleons. We explore the
  analytical structure of the equation in the complex $k$-plane by
  means of Jost functions. We find that despite the Mott effect the
  correlation with deuteron quantum numbers are manifested as
  antibound states, i.e., as zeros of the Jost function on the
  negative imaginary axis of the complex momentum space.  The analysis
  presented here is also suited for Coulombic systems.


{$^*$Permanent address:
Fachbereich Physik, Universit\"at Rostock, 18051 Rostock,
Germany}\\

\end{abstract} 
\vspace{5mm}

]

\section{Introduction}

Nuclear matter exhibits quite a rich phase structure because of strong
correlations between nucleons. The formation of clusters below a
certain density (Mott density) and the appearance of superconductivity
below a critical temperature are consequences of these correlations.
An understanding of nuclear matter at finite temperatures, e.g., the
knowledge of the equations of state, is largely relevant for a
description of heavy-ion reactions and astrophysical objects such as
the formation and structure of neutron stars.

To describe correlations requires a treatment that goes beyond the
simple picture of noninteracting quasiparticles in a background field.
A powerful formalism for a systematic treatment of correlations is
provided by the Dyson equations approach, for a review see
Ref.~\cite{duk98}.  This approach has been used to derive effective
in-medium equations that can be solved rigorously with few-body
techniques~\cite{Beyer:1996rx,Beyer:1997sf,Beyer:1999tm,Beyer:1999zx,Kuhrts:2000jz,Beyer:1999xv,Beyer:2000ds,Kuhrts:2001zs,alt67,san74,alt72,san75}.
For two-body correlations it leads to equations known as
Galitskii-Feynman or Bethe-Goldstone
equations~\cite{fet71,schmidt:1990,schnell:1996,stein:1995,Bozek:1999su}.
These equations include the dominant effects of the medium,
i.e.  the self-energy corrections and the Pauli blocking.  Since the
residual interaction depends on the Pauli blocking factors the
effective potential becomes explicitly momentum dependent.  Therefore
the effective equations so far have been solved in momentum space
only.

Here we present a solution of the effective in-medium two-body
equation in co-ordinate space and focus on the analytical structure of
the residual interaction between quasiparticles.  Although, in
general, solutions should not depend on the chosen representation, it
is instructive to construct an equivalent effective local potential.
In addition, the techniques presented here are not only suited for the
two-nucleon problem, but also applicable for Coulombic systems where
in the simplest case the effective interactions are given in $r$-space
supplemented by a Debye screening.

The interesting region of the phase diagram is that close to the Mott
transition which indicates a transition from a gas of nucleons to a
gas of nuclids.  Below the Mott transition correlations arise as bound
states, i.e. a pole can be identified for energies $E(\mu,T,P)<0$. The
energy depends on the chemical potential $\mu$, the temperature $T$
and the c.m. momentum $P$ of the cluster. Above the Mott transition
bound states ``disappear'', however, correlations are still present
and can be found by solving the respective scattering problem. We
explicitely show that for the two-nucleon problem, to begin with, a
deuteron-like in-medium state can be identified as an anti-bound
state.  To arrive at this result we use the technique of Jost
functions elaborated for local potentials in
\cite{res1,res2,res3,mass}. The application to larger clusters will be
postponed to a future work.

In Sec.~\ref{sec:form} we describe the formalism that defines the
in-medium equations and the resulting medium-modified effective
nucleon-nucleon potentials for quasiparticles. We also briefly recall
the Marchenko inversion method to derive phase equivalent local
potentials and the exact method of obtaining potential resonances.
Our results are presented in Sec. III and our conclusions in Sec. IV.

\section{Formalism}\label{sec:form}

\subsection{Effective in-medium Equations}
\label{sec:IME}
Within the framework of thermodynamic Green functions effective
in-medium two-body equations known as Galitskii-Feynman or
Bethe-Goldstone equations (depending on details) can be
derived~\cite{fet71}.  The residual potential that enters into the
effective equations, given below, includes Pauli blocking factors.  The
Schr\"odinger type in-medium equation used here is given by
\begin{equation}
(H_0  - z) \Psi(12)  +
\sum_{1'2'} (1-f_1-f_2) V_2(12,1'2') \Psi(1'2') = 0.
\label{eqn:S}
\end{equation}
where the bare two-nucleon potential is denoted by $V_2$. The
Matsubara frequency $z_\lambda=\pi\lambda/(-i\beta)+2\mu$ where $\mu$
is the chemical potential and $\gb=1/k_BT$ the inverse temperature has
been analytically continued, i.e $z_\lambda\rightarrow
z$~\cite{fet71}. The Fermi function is
\begin{equation}
f_1\equiv f(k_1) = \frac{1}{e^{\gb(\gve(k_1) - \mu)}+1}
\label{eqn:Fermi}
\end{equation}
The single-nucleon density $\rho(\mu,T)$ can be calculated in the
standard manner~\cite{fet71}. The quasiparticle self energy $\gve$
that appears in Eq.  (\ref{eqn:Fermi}) solves the one particle Green
function. For an uncorrelated medium the Hartree-Fock approximation
results in
\begin{equation}
\gve(k_1)=\frac{k_1^2}{2m_1}+\sum_{2}V_2(12,\widetilde{12})f_2
\equiv\frac{k_1^2}{2m_1}+\gD^{\rm HF}(k_1),
\label{eqn:eps}
\end{equation}
where $\widetilde{12}$ denotes proper antisymmetrization of particles
1 and 2. Hence the effective Hamiltonian of noninteracting identical
quasiparticles that enters into eq.~(\ref{eqn:S}) is given by
\begin{equation}
H_0=\sum_{i=1}^2  \left(\frac{k_i^2}{2m}+\gD^{\rm HF}(k_i)\right).
\label{eqn:H0}
\end{equation}

For the moment we are interested only in a region of rather low
density in the vicinity of the Mott transition ($\sim \rho_0/10$,
where $\rho_0=0.16$ fm$^{-3}$ is normal nuclear matter density). We
therefore use an effective mass approximation for the quasiparticle
energy $\gve$; i.e. after evaluation of Eq.~(\ref{eqn:eps}) we fit the
effective mass $m_{\mathrm{eff}}$ via
\begin{equation}
\gve=\frac{k^2}{2m}+\gD^{\rm HF}(k)
\simeq\frac{k^2}{2m_{\mathrm{eff}}}+\gD^{\rm HF}_0.
\end{equation}
The constant shift $\gD^{\rm HF}_0$ can be absorbed in a redefinition
of the chemical potential $\mu_{\rm eff}=\mu-\gD^{\rm HF}_0$.  Further
to simplify our analysis we consider the two-body system to rest in
the medium. It is well known that the influence of the medium fades
for larger relative momentum respective to the medium
\cite{schmidt:1990}.  Upon introducing relative and c.m.  coordinates,
$p$ and $P=0$, for the two-particle system the kinetic energy term
reads
\begin{equation}
\frac{k_1^2}{2m_{\mathrm{eff}}}=\frac{k_2^2}{2m_{\mathrm{eff}}}
=\frac{p^2}{2m_{\mathrm{eff}}}=\frac{1}{2}\, E
\end{equation}
where $E=z-E_{\mathrm{cont}}$ is the two-body binding/scattering energy in the
c.m. system; the continuum edge for this case $P=0$ and effective mass
approximation is given by $E_{\mathrm{cont}}=2\gD^{\rm
  HF}_0$\cite{schmidt:1990}.  Using a local bare nucleon nucleon
potential $V_2(r)$, where $r$ is the conjugate of $p$, the
effective potential becomes energy dependent
\begin{equation}
          V(E,r)=(1-2f(E))\;V_2(r), 
\label{vef}
\end{equation}
where the Fermi function is now explicitly given by
\begin{equation}
f(E)=\frac{1}{e^{\gb(E/2 - \mu_{\mathrm{eff}})}+1}.
\end{equation}
The effective
in-medium Schr\"odinger-type equation for scattering states then reads
\begin{equation}
          \left(\frac{p^2}{m_{\mathrm{eff}}} + V(E,r) - E\right)\;\psi(E,r) = 0.
\end{equation}
Note that this equation holds for $E>0$, since
$E=p^2/m_{\mathrm{eff}}$ for the eigenvalues.

The strategy is as follows. We study the question of what happens to
the cluster (for the time being the deuteron) for densities above the
Mott density and explore the location of possible resonances. One may
is to solve the equations on the unphysical energy sheet (second
sheet); for a textbook treatment see, e.g. \cite{gloeckle:1988}.
Alternatively, we way utilize the properties of Jost functions as
elaborated in \cite{res1,res2,res3,mass}.  To do so we first derive a
phase equivalent local potential to Eq. (\ref{vef}) for each chemical
potential $\mu$ and temperature $T$ of interest.  This will be
explained in the next subsection \ref{sec:ISM}.  The next step is to
construct the appropriate Jost function and explore the analytical
structure by analytic continuation of $r$ to the complex plane as
explained in the subsection \ref{sec:APJ}. The zeros of the Jost
functions in the complex ``$k$-plane'' are related to bound states,
resonances, or anti-bound states that can all be identified.

\subsection{Inverse Scattering Method}
\label{sec:ISM}

A way to construct the nucleon-nucleus potential in configuration
space is to use inverse scattering techniques. The potential in this
case is directly obtained from the available scattering information
and the bound states.  In the present work we shall employ the inverse
scattering method at a specific partial wave $\ell$ (fixed-$\ell$
inversion) which has been the subject of several books and monographs
(see for example, Refs. \cite{Chadan,Newton,Mar}).  Therefore,
we shall outline here only the main features and formulas for
convenience.

Most applications at fixed angular
momentum employ the so-called exactly solvable classes of potentials
of Bargmann type \cite{Barg}. In this approach the $S$-function,
defined as a ratio of Jost functions,
\begin{equation}
        S_\ell(k)=\frac{f_\ell(-k)}{f_\ell(k)}\,,
\label{Sk0}
\end{equation}
is parameterized using the rational form 
\begin{equation}
          S_\ell(k)=\prod_{n=1}^{N_b} \frac{k+ ib_n}{k- ib_n}\ 
          \prod_{m=1}^{M} \frac{k+ a_m}{k- a_m}\ ,
\label{rat}
\end{equation}
where $N_b$ is the number of bound states and $M$ the number of terms
needed to reproduce the phase shifts from 0 to $\infty$. This
parametrization ensures the existence of bound states at $k=ib_n$ and
leads to a special class of potentials known as Bargmann potentials
(see Refs.~\cite{Chadan,Newton} for more details). They lead to
kernels $K_\ell(r,r')$ of the exact Marchenko \cite{Mar} and
Gel'fand-Levitan \cite{Lev} integral equations from which the
potentials can be constructed. The method has been successfully
applied to construct nucleon-nucleon potential by von Geramb and
collaborators (see Refs. \cite{Ger,Apa} and references therein), as
well as neutron-deuteron potentials \cite{Pap,Alt}.

The Marchenko integral equation, fully  equivalent to the
Schr\"odinger equation, reads
\begin{equation}
        K_\ell(r,r')+{\cal F}_\ell(r,r')+\int_r^\infty
                K_\ell(r,s){\cal F}_\ell(s,r'){\rm d} s=0\ .
\label{March}
\end{equation}
The kernel  ${\cal F}_\ell(r,r')$ of this equation is related to the
$S$-matrix $S_\ell(k)$, and thus to experiment,  via
\begin{eqnarray}
     {\cal F}_\ell(r,r')&=&\frac{1}{2\pi} \int_{-\infty}^\infty
     h_\ell^{(+)}(kr)\left[1-S_\ell(k)\right]h^{(+)}_\ell(kr')
      {\rm d}k\nonumber\\&&+
      \sum_{n=1}^{N_b}
     A_{n \ell}h^{(+)}_\ell(b_n r)h^{(+)}_\ell(b_n r')\ ,
\label{frr}
\end{eqnarray}
where $h^{(+)}_\ell(z)$ is the Riccati-Hankel function, $N_b$ is the
number of bound states, and $A_{n \ell}$ are the corresponding
asymptotic normalization constants (in what follows we will assume
that only one bound state is present in the system).  Once the Fourier
transformation, Eq. (\ref{frr}), is obtained, the potential
$V_\ell(r)$ for the partial wave $\ell$ is calculated from the
relation
\begin{equation}
        V_\ell(r)=-2\;\frac{{\rm d} K_\ell(r,r)}{{\rm d}r}.
\label{Vr}
\end{equation}
With the choice (\ref{rat}) and for $N_b=1$ the
integration in (\ref{frr}) can be readily obtained,
\begin{eqnarray}
     {\cal F}_\ell(r,r')&=&-i\sum_{m=1}^{M_u} {\cal R}_m^\ell 
        w^+_\ell(\alpha^\ell r) w^+_\ell(\alpha^\ell r')
\nonumber\\
&&       +A_\ell w^+_\ell(ib_l r) w^+_\ell(ib_\ell r'),
\label{Frr}
\end{eqnarray}
where ${\cal R}_m^\ell$ are the coefficients from the 
residues of the integrand for the $M_u$ poles of  (\ref{rat})
lying in the upper complex $k$-plane (excluding the 
one corresponding to the bound state $ib_\ell$).

Two aspects of the Marchenko scheme should be emphasized.  First, in
the absence of bound states the method provides us with a unique,
shape-independent (in the sense that the shape is not prechosen) but
$\ell$-dependent interaction that reproduces the phase-shifts at all
energies. Second, in the presence of bound states the potential is not
unique as one requires the knowledge of the asymptotic normalization
constant which can be chosen to have any arbitrary value. This will
result in a set of equivalent local potentials all reproducing the
scattering data equally well.  However, the choice
\begin{equation}
        A_{\ell}=i\frac{f_\ell(-ib)}{\dot f_\ell(ib)},
\label{asym}
\end{equation}
where $\dot f_\ell(ib)={\rm d}f_\ell(k)/{\rm d} k$ provide us with a
short-range unique potential, see Ref.~\cite{Chadan} for a discussion.
We mention here that once the rational fit (\ref{rat}) is achieved the
extraction of $f_\ell(k)$ is obtained from the procedure described by
Massen et al. \cite{mass}

In the present work the potentials obtained are energy dependent and
therefore the bound state for the medium modified interaction is not
known. However, the fit (\ref{rat}) for the cases where the phase
shifts amount to $\pi$ at the origin -- irrespective of the number of
terms used -- always provides a constant pole on the positive
imaginary $k$-axis and when ignored (i.e. when it was considered as a
pole of $f_\ell (-k)$ and thus spurious), highly repulsive potentials
behaving like $\sim 1/r^2$ for small $r$ were obtained. These are
nothing else but the shallow supersymmetric partner potentials
\cite{mass,baye}. In contrast, when interpreted as a true binding
energy with the corresponding asymptotic normalization constant being
calculated via Eq. (\ref{asym}), the fit provided us with a short range
potential reproducing the same scattering data of the energy dependent
effective potential, Eq. (\ref{vef}).

\subsection{Analytic Properties of the Jost function}
\label{sec:APJ}
Extraction of the rational parameterization of the Jost function does
not always provide us with the physical poles as some of them may be
an artifact of the fitting procedure. To extract the actual potential
resonances one may apply one of the so-called complex energy methods.
In Refs. \cite{res1,res2,res3,mass} a new method has been suggested in
obtaining the analytical properties of the Jost function in the whole
complex $k$-plane.  This method consist of replacing the Schr\"odinger
equation by an equivalent system of linear first-order differential
equations.  By solving this system for real momenta $k$ one obtains
the Jost solutions and therefore the Jost function from which the
phase shifts can be calculated.  Solutions for which the Jost function
has a zero on the imaginary axis provide us the {\em bound} and {\em
  anti-bound} states (zeros on the positive and negative imaginary
axis respectively), while by locating a complex zero $k_r$ of the Jost
function in the fourth quadrant of the momentum plane, one obtains the
{\em resonances} of the underlying interaction.

In order to calculate the Jost function for any value of momentum
on the complex $k$-plane we perform a complex rotation of the
coordinate $r$
$$
  r=x\exp(i\phi)\qquad x\ge0\qquad \phi\in [0,\phi_{\rm max}], 
   \qquad  \phi_{\rm max}<\pi/2
$$     
in the Schr\"odinger equation and look for a solution of 
the form
\begin{equation}
\label{ansatz}
     \phi_\ell(k,r)=\frac{1}{2}\left[
                  h^{(+)}_\ell(kr)F^{(+)}_\ell(k,r)
                  +h^{(-)}_\ell(kr)F^{(-)}_\ell(k,r) \right]
\end{equation}
where $  h^{(\pm )}(kr)$ are the Riccati-Hankel functions
and $F^{(\pm)}_\ell(k,r)$ are auxiliary functions fulfilling
the first order differential equations \cite{res1}
\begin{eqnarray}
\nonumber
\partial_rF_\ell^{(+)}(k,r)&=&\phantom{+}\displaystyle
          \frac{h_\ell^{(-)}(kr)}{2ik}
          V_\ell(r)\left[
          h_{\ell}^{(+)}(kr)F_\ell^{(+)}(k,r)\right.\\\nonumber&&\left.+
          h_{\ell}^{(-)}(kr)F_\ell^{(-)}(k,r)\right]\ ,\\
\label{fpmeq}
&&\\
\nonumber
\partial_rF_\ell^{(-)}(k,r)&=&-\displaystyle
          \frac{h_\ell^{(+)}(kr)}{2ik}
          V_\ell(r)\left[
          h_{\ell}^{(+)}(kr)F_\ell^{(+)}(k,r)\right.\\\nonumber&&\left.+
          h_{\ell}^{(-)}(kr)F_\ell^{(-)}(k,r)\right].
\end{eqnarray}
From the
solution of this system   one obtains 
the regular solution $\phi_\ell(k,r)$ 
The use of the system (\ref{fpmeq}) 
has the main advantage that it
is exact with rigorous boundary conditions,
\begin{equation}
\label{assansatz}
          \phi_\ell(k,r)\mathop{\longrightarrow}_{r\to\infty}\frac12\left[
          h_{\ell}^{(+)}(kr)f_\ell^{*}(k^*)+
          h_{\ell}^{(-)}(kr)f_\ell(k)\right],
\end{equation}
and that  
the function $F^{(-)}_\ell$ coincides  asymptotically with the
Jost function, viz.
\begin{equation}
\label{Jf}
         F_{\ell}^{(-)}(kr)  \mathop{\longrightarrow}_{r\to\infty}
         f_\ell(k)
\end{equation}
More details on the method can be found in Refs. 
\cite{res1,res2,res3,mass}.
%

\section{Results}
\begin{figure}
\begin{center}
\psfig{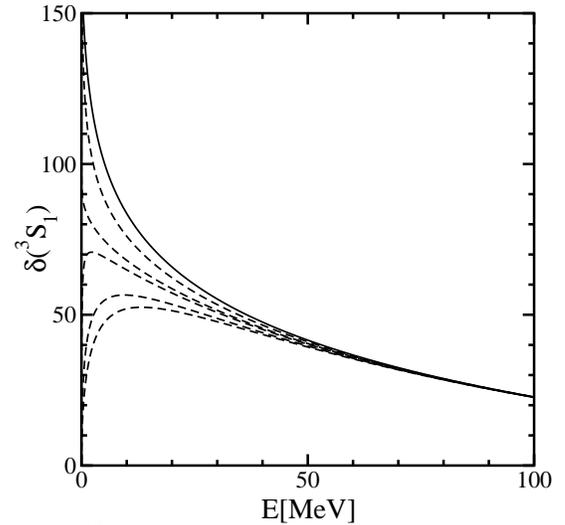}
\caption{\label{fig:phase}
  $^3$S$_1$ phase shifts for the potential at five different densities
  (dashed lines).  Higher densities deviate stronger from the isolated
  MT phase-shifts (solid line). For densities 0.009, 0.017, and 0.034
  fm$^{-3}$ the phase-shifts start from zero implying that the
  potential does not sustain a bound state.}
\end{center}
\end{figure}

In the present study we employ the central part of the Malfliet-Tjon
I+III (MT) potential \cite{MT}, which is given in $r$-space, as the
bare potential $V_2$.  The three-nucleon binding energy for this
potential is 8.595\,MeV,  i.e.  very close to the experimental value
of 8.48\,MeV. The medium-modified energy dependent potentials $V(E,r)$
for five different densities were constructed using Eq. (\ref{vef}).
The in-medium parameters for these potentials are given in Table
\ref{tab:TDE}. 
\begin{table}[bht]
\caption{\label{tab:TDE}
The potentials used for $T=10$\,MeV and the resulting poles on the
imaginary axis of the $k$-plane. The plus sign for the pole corresponds
to bound and the minus to anti-bound states. (For the meaning of
symbols see text.)}
\begin{tabular}{lllll}
 $V(E,r)$  & $\rho[{\mathrm fm}^{-3}]$    & $\mu_{\mathrm{eff}}$ [MeV]   
       & $m_{\mathrm{eff}}$ [MeV]  & Pole $ib$ [fm$^{-1}$] \cr
\hline
   MT  & $ -$    &   $ -$    &939       &$ +0.2324i$\cr
   V003&0.003  &$-22.50$  &925.43  & $+0.1132i$\cr
   V007&0.007  &$-13.29$  &907.99  & $+0.0055i$\cr
   V009&0.009  &$-10.39$  &899.69  & $-0.0353i$\cr 
   V017&0.017  &$-2.488$   &868.04  & $-0.1680i$\cr 
   V034&0.034  &$+7.909$   &809.65  & $-0.2640i$
\end{tabular}
\end{table}
\begin{figure}
\begin{center}
\psfig{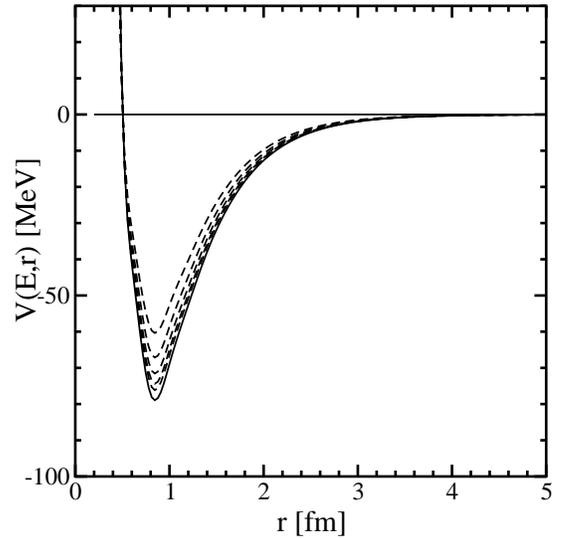}
\caption{\label{fig:vve_009}
  Energy dependent potentials for $T=10$\,MeV and density $n=$ 0.009
  fm$^{-3}$ (dashed lines) for energies $E=50$, 40, 30, 20, and 10
  MeV. Lower energies deviate stronger from the isolated MT potential
  (solid line).}
\end{center}
\end{figure}
 These are the effective chemical potential
$\mu_{\mathrm{eff}}$ and the effective mass $m_{\mathrm{eff}}$. For
simplicity we have used values calculated und used earlier
\cite{Beyer:1996rx}.  
In Fig.  \ref{fig:vve_009} we present the
potentials $V(E,r)$ for a nuclear density $\rho=0.009$ fm$^{-3}$ and
for energies $E=10, 20, 30, 40$, and $50$\,MeV and compare them to the
original MT potential. It is seen that the potentials become shallower
but eventually at higher energies the in-medium effects are minimal
and the MT is practically recovered.  
The corresponding phase shifts
for these potentials are shown in Fig.  \ref{fig:phase}.  
\begin{figure}
\begin{center}
\psfig{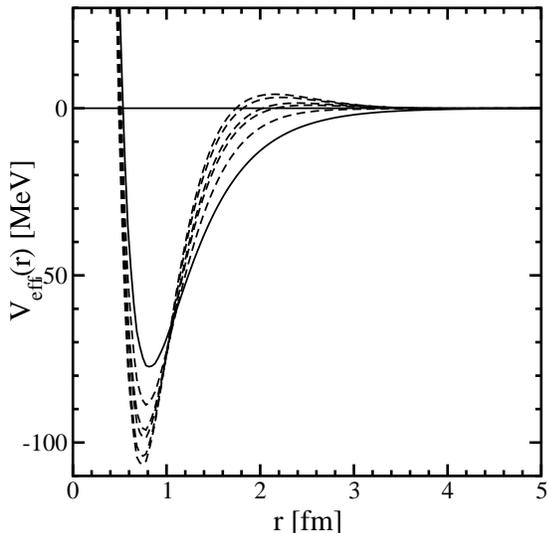}
\caption{\label{fig:vvm}
  The equivalent local potentials $V_{\mathrm eff}(r)$ obtained via
  Marchenko inversion method for five densities at $T=10$ MeV (dashed
  lines). Higher densities deviate stronger from the isolated MT
  potential (solid line).}
\end{center}
\end{figure}
Note that
the phase-shifts for the densities 0.009, 0.017, and 0.034 fm$^{-3}$
start from zero implying that the underlying potential does not
sustain a bound state. This is in agreement with the results of Ref.
\cite{schmidt:1990}, where a separable version of the Paris potential
has been used as a bare nucleon nucleon potential.

The energy-dependent potentials as given in Eq. (\ref{vef}), albeit
more realistic than the energy independent ones, cannot provide us
with the bound state or the resonance structure of the interaction.
To this end we need to construct an energy independent equivalent local
potential that can be analytically continued to complex $r\rightarrow
r{\rm e}^{i\varphi}$ values, with $0\le\varphi\le\pi/4$ and therefore
the analytical properties of the Jost function in the whole $k$-plane
can be explored.

To construct energy independent potentials we employ the Marchenko
inverse scattering method. As discussed in Sec. \ref{sec:form} this
method requires the phase-shifts at all energies. The quality of these
potentials depends on the cut-off energy $E_{\rm max}$.  For
sufficiently high $E_{\rm max}$ (in our case $\sim$1500\,MeV) the
off-shell differences of the equivalent potentials, i.e. the
differences of the scattering function in the interior region are very
small. The resulting potentials are presented in Fig.  \ref{fig:vvm}.
Two essential features of these interactions should be noted. The
attractive well becomes deeper but narrower while at the same time a
repulsion in the interaction region appears as the density of the
medium increases. This amounts to an decrease in the binding energy
until, at the Mott density, the two nucleons become unbound.  Due to
the repulsion in the interaction region, although quite small, such a
potential is expected to give rise to resonances. We search for
resonances in the fourth quadrant of the $k$-plane as explained in
Sec.\ref{sec:APJ}. We find that the bound state, instead of being
continued off the imaginary axis, moves along 
\begin{figure}
\begin{center}
\psfig{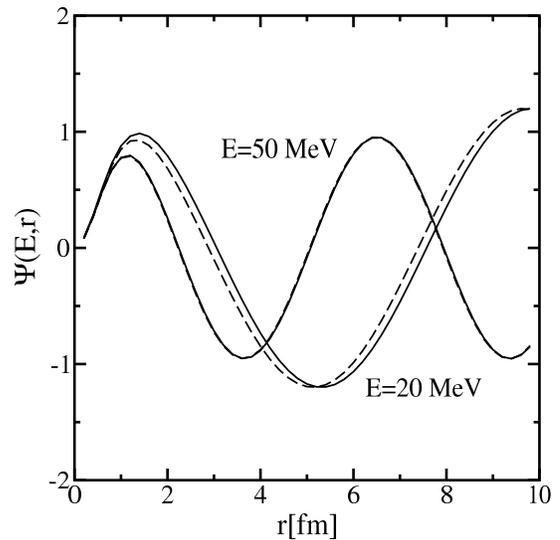}
\caption{\label{fig:wave}
  Comparison of the scattering wave functions $\Psi(E,r)$ obtained
  with the $E$-independent Marchenko potential (dashed lines) with
  those of $V(E,r)$ (solid lines) for $n=$0.007 fm$^{-3}$ at $E=20$
  and 50 MeV.}
\end{center}
\end{figure}
\noindent the negative imaginary
$k$-axis, i.e.  it becomes an {\em anti-bound} state.  Numerical
results are given in Table \ref{tab:TDE} where $ib$ denotes the
respective position of the bound ($b>0$) or anti-bound ($b<0$) state.

The quality of the potential obtained via inversion is demonstrated in
Fig. \ref{fig:wave} where the scattering wave function generated by
the energy independent Marchenko potential $V_{\rm eff}(r)$ for
$n=$0.007 fm$^{-3}$ and at $E=20$ and 50\,MeV, is compared with the
respective wave function for $V(E,r)$ of Eq. (\ref{vef}).  The
agreement of the respective two wave functions is remarkable.  Beyond
the interaction region, the two wave functions are, to all practical
purposes, identical. Similar results were obtained at all energies and
potentials.

The potentials obtained via inversion are given in numerical form.
Therefore we follow here the same procedure as used by Massen et al
\cite{mass}, viz. parameterize the potentials using analytical {\em
  ansatzes}, with parameters obtained via the MERLIN minimization
program \cite{MERLIN}. As the potentials are intended for use in
AGS-type equations, we use for this purpose a sum of Yukawa terms and
the minimization was terminated once an accuracy of better than $\sim
10^{-5}$ was achieved at all points.


\section{Conclusions}

In the present work we constructed ener\-gy-in\-de\-pen\-dent but
$\ell$-dependent potentials equivalent to the medium modified
interaction. The purpose of this endeavor is multifold. First, to
construct potentials for interpretation purposes and to find how the
medium modifications are manifested in $r$-space. Secondly, to
generate the necessary numerical methods that can readily be used to
construct energy-independent potentials suitable for three- and
four-nucleon in-medium calculations.  Thirdly, to find the analytical
properties of the corresponding two-body Jost function -- the first
step towards the study of resonances in nuclear medium using the
method described in section IIC.  Last but not least, we argue that
this method although applied in a nuclear physics context, can be
utilized in other strongly correlated Fermi systems.

Two specific features of the constructed interactions should be
emphasized here; the first one is the movement of the attractive well
towards the interior region while at the same time is becomes narrower
as the densities become smaller. The second is the appearance of
anti-bound states instead of resonances in first quarter as the
density increases.  This reflects a vanishing imaginary part of the
respective two-body spectral function.  Hence, the two-body cluster
(even for $b<0$) retains the character of a quasi-particle, i.e. the
correlation does not decay. In a more general approach the
self-consistency requirement~\cite{Bozek:1999su} as well as three-body
collisions~\cite{Beyer:1997sf} lead to imaginary parts of the two-body
spectral function, and therefore to a finite width (life time) of the
two-particle cluster irrespective of $b$ smaller or larger than zero.

It remains to be seen how these modifications are
manifested in the three- and four-body systems embedded in the
nuclear medium especially in nuclear scattering, in photo- and
electro-processes and in the appearance or otherwise of resonances in
the relevant clusters. 

\section{Acknowledgments}
We thank G. R\"opke for discussions. This work is supported by
Deutsche Forschungsgemeinschaft BE 1902/7.



\end{document}